\documentclass[aps,prd,twocolumn,superscriptaddress,groupedaddress,showpacs]{revtex4-1}
\usepackage{empheq}
\usepackage{amsmath}
\usepackage{framed, multido}
\usepackage{graphicx}
\usepackage{dcolumn}
\usepackage{bm}
\usepackage[mathlines]{lineno}

\usepackage{tensor}
\usepackage{mathtools}
\usepackage{natbib}
\usepackage{enumerate}
\usepackage{comment}
\usepackage{amsfonts,amssymb,amsthm,array,amsmath,mathrsfs}
\usepackage{relsize}
\usepackage{cancel}
\usepackage{mathtools}
\usepackage[T1]{fontenc}
\usepackage{fancyhdr}

\usepackage{soul}

\usepackage[english]{babel}
\usepackage[utf8]{inputenc}

\usepackage[normalem]{ulem}

\usepackage{subfigure}

\newcommand{\dd}{{\rm{d}}}

\newcommand{\eq}[1]{(\ref{#1})}

\newcommand{\la}{\label}
\newcommand{\ba}{\begin{align}}
\newcommand{\ee}{\end{equation}}
\newcommand{\be}{\begin{equation}}

\def\12{\frac{1}{2}}
\newcommand{\p}{\partial}
\newcommand{\en}{\end{align}}
\newcommand{\e}{\epsilon}

\usepackage{color}

\begin{document}

\title{Multivalued Wess-Zumino-Novikov Functional and Chiral Anomaly in Hydrodynamics}
\author{P.B.~Wiegmann}
\affiliation{
Kadanoff Center for Theoretical Physics, University of Chicago,
5640 South Ellis Ave, Chicago, IL 60637, USA
}


\date{\today}





\begin{abstract}
We present a hydrodynamic framework derived from the action of a perfect fluid, modified by the hydrodynamic analog of Novikov's multivalued functional. This modification introduces spin degrees of freedom into the fluid. The structure closely resembles the Abelian version of the Wess-Zumino functional, commonly applied in field theories with chiral anomalies. The deformation incorporates transport properties of Weyl fermions and, in the case of a charged fluid, exhibits the chiral anomaly. It is also consistent with Onsager's semiclassical quantization of circulation. Additionally, we discuss the hydrodynamic analog of instantons and related topological invariants.

\end{abstract}
\maketitle

\section{Introduction and the background}\la{Sec1}
The equations of motion  for a perfect fluid can be regarded as 
conservation laws associated with the group of spacetime diffeomorphisms. If no external forces act on the fluid the  momentum and energy are conserved resulting in four dynamical equations  expressed as a divergence-free condition of the canonical  momentum-stress-energy tensor \begin{align}
\p_\mu\tensor{T}{^\mu_\nu}=0\,.\la{T}
\end{align} These equations
alone are sufficient for describing
  flows, when the only dynamical variables are components of   the particle 
number 4-current  \(n^\mu=(n^0,\,
n^0\bm v)\), where \(n^0\) is the particle number density and \(\bm v\) is the fluid velocity. In this case the continuity equation   
 \begin{align}
\p_\mu n^\mu=0\,.\label{153}
\end{align}  is not an independent condition. It follows from the conservation of momentum and energy described by \eq{T}. Barotropic flows and more general homentropic flows represent this situation. 

A more general, baroclinic flow  involves
additional dynamical variables. In this case, more equations are necessary. They stem from symmetries other than spacetime diffeomorphisms, such as gauge symmetry. Then the continuity equation \eq{153} normally arises as the Noether conservation law for a particle number.

The phenomenon known as the {\it chiral current anomaly} presents an obstacle
to the conservation of particle number. The issue arises when the conserved Noether current generated by the gauge symmetry, denoted as \(I^\mu\),   is  not gauge-invariant;
however, its divergence is.    In this case, the particle number  is not identical to the Noether charge,  and is not conserved \(\p_\mu n^\mu\neq 0\). Such system is not isolated  and in contact with a reservoir capable to supply or remove particles. At the same time the equations of motion {   } 
\begin{align}
\p_\mu I^\mu=0\,\label{224}
\end{align} remains local and gauge-invariant. The chiral anomaly signifies that the flow entrains a  reservoir capable
 of supplying and swapping particles. 

The chiral anomaly was initially identified as a kinematic property of quantum field theories
involving chiral (or Weyl) fermions \cite{Current_algebra}.  A defining feature of the chiral anomaly is that the particle production rate,
\(\p_\mu n^\mu\), is locally defined by the flow itself and is unaffected by changes in the spacetime metric.
Therefore, the anomaly is  largely insensitive
to interaction and, when carried over to a liquid state
 it does not
introduce additional spacetime scales beyond already accounted gradients of hydrodynamic fields. Being insensitive to a variation of metric, the chiral anomaly only impacts the continuity equation while leaving the form of the stress  tensor and its conservation \eq{T}  unaffected. 

In recent years, there has been
growing confidence that the current  anomalies
 are compatible with classical fluid dynamics. An incomplete list of references is \cite{volovik, Erdmenger,Banerjee2011,son2009hydrodynamics,stephanov2012chiral,
 jensen2013thermodynamics,haehl2014effective,
stephanov_son2015,yamamoto,Kharzeev,AMN,abanov2022axial,wiegmann2022chiral,AW2,wiegmann2023Hamilton,
Abanov_Cappelli}. More references could be found in the review \cite{Y}.
A physical argument supporting this perspective is the existence of liquids composed of Weyl fermions. Such liquids are expected to retain  kinematic features of Weyl fermions, including their anomalies.
Notable examples  are  the superfluid \(^3 {\rm He\, A}\), {  semiconductors with high spin-orbit interaction, and quark-gluon plasma occurring in  heavy-ion collisions (see e.g., \cite{volovik,Landsteiner,Kharzeev,Y} for review of each topic).}

The gauge symmetry generated Noether current is expected to be equal a particle number
current \(n^\mu\) modified by a  pseudovector field \(h^\mu\)
  \begin{align}
I^\mu=n^\mu+ \tfrac k 2h^\mu\,.\label{183}
\end{align}  
 Here 
\(k\), referred as the {\it level},  is a pseudoscalar parameter representing the strength of the deformation.      Then the Noether conservation law \eq{224}   becomes the   equation for the particle production 
 \begin{align}
\p_\mu n^\mu=- \tfrac k 2\p_\mu h^\mu\,.\la{51}
\end{align}
The imposed properties on \(h^\mu\) are as follows: it is (i)  locally expressed through
Eulerian fields, (ii)  does not possess any scale, and therefore has no reference to the spacetime metric, (iii) its divergent 
\(\p_\mu h^\mu\)    is    gauge-invariant. 

In fluid mechanics there is only one object of this kind,  
 the  4-{\it fluid helicity} current.   The fluid helicity
is the dual to the 3-differential form \(h=p\wedge dp\) constructed from
of fluid   4-momentum 1-form  \(p=p_\mu dx^\mu\). In tensor
notations \begin{align}
    h^\mu=\varepsilon^{\mu\nu\lambda\sigma}p_\nu\p_\lambda
p_\sigma\,,\label{1741}
\end{align}
where \(\varepsilon^{\mu\nu\lambda\sigma}\) is the Levi-Civita symbol. The  time-like component of the helicity current \(h_0={
\bm
p}\cdot\nabla\times \bm p\) is  the usual helicity density
[in Sec.IV we give a formal definition of the  kinematic
momentum and its relation to the particle number current]. 
Variants of such   deformation, albeit in various different settings   had been introduced in 
Ref. \cite{Erdmenger,Banerjee2011,son2009hydrodynamics, abanov2022axial}. An incomplete list of early   related works  is  \cite{
jensen2013thermodynamics, haehl2014effective, stephanov_son2015, yamamoto}.

We  remark    
that a room for 
 deformations of fluid dynamics is limited, as the fluid equations of motion must be covariant
under the  action of the gauge  group and   the group
of spacetime  diffeomorphisms
\be\mathcal{G}=U(1)\rtimes{\rm\ Diff}(\mathcal{M}^4)\,. \la{G}\ee 

Why would the deformation  \eq{51} be consistent with fluid dynamics?
A concise criterion is the  Hamilton
principle of fluid mechanics. The Hamilton principle asserts the existence
of a Hamilton functional or an action, whose invariance under the action of the
symmetry group \(\cal G\) yields the desired  equations of motion.
This is the central part  of our approach.
We construct the Hamiltonian functional  which yields the Eqs. \eq{T} and \eq{51} keeping   the stress tensor of the perfect fluid  intact. The latter is given by \begin{align}
\tensor{T}{^\mu_\nu}=n^\mu p_\nu+\tensor{\delta}{^\mu_\nu}P\,.\label{T1}
\end{align}
Here \(P\) is the fluid pressure. Being complemented by the equation of state which expresses the fluid momentum in terms of the particle current (discussed  below) Eqs. (\ref{T},\ref{51},\ref{T1}) give the complete  set of the  fluid equations of motions.

We will show that the  action or the Hamiltonian functional of the perfect fluid could be uniquely extended by the
multivalued Wess-Zumino-Novikov (WZN) functional and that this extension yields the  Eqs. (\ref{T},\ref{51},\ref{T1}).
The multivaluedness of the WZN action provides global obstructions which restricts the parameter
   \(k\) to  be  
an integer in units of the Planck constant. This result is  consistent with the Onsager quantization and the known kinematic properties of Weyl fermions developed in the early works
of Vilenkin \cite{Vilenkin}.
Our results are summarized by (\ref{46},\ref{47},\ref{48}) in Sec.\ref{Sec10}. 

Our construct holds for any even spacetime dimension \(d\). In this case,
 the modification  of the Noether current  \eq{183} is given by the  \((d-1)\)-form
\begin{align}
\frac k{(d/2)!}p\wedge(d p)^{d/2-1}\,.
\end{align}
In particular,  this formula agrees with
 the known expression for the Noether current of  (1+1)-chiral bosons \be
I^\mu=n^\mu+k\varepsilon^{\mu\nu}p_\nu\,.\ee

      The concept of the multivalued functionals in a general setting was introduced  by Novikov in 1981 \cite{Novikov}. Soon after Novikov's paper, it was recognized that a class of these functionals  appeared in the early work of Wess and Zumino \cite{Wess-Zumino}. Wess and Zumino constructed the functional whose variation replicates effects of the anomaly. In this paper we introduce the hydrodynamic version of the  multivalued functional.

 The anomaly is a topological phenomenon in the sense that it is metric-independent and therefore could be expressed solely in terms of differential forms.
The efficient framework that helps incorporate anomalies into fluid mechanics is the spacetime covariant formulation of hydrodynamics of Lichnerowicz \cite{Lichnerowicz} and Carter \cite{Carter}. For recent reviews, see \cite{Gourgoulhon, Markakis}, and \cite{AMN, AW2, wiegmann2023Hamilton} for its adaptations to anomalies. In this approach, the hydrodynamics is expressed in terms of   the particle number 4-current 
\(n^\mu\), and its conjugate,  a covector, the  fluid 4-momentum \(p_\mu\),  without reference to the
spacetime metric. Consequently, the form of the fluid equations of motion  appear identical for both relativistic and non-relativistic fluids. 

We begin with a discussion of the relation between Eq.\ref{51} and the traditional form of the anomaly as a linear response to an external electromagnetic field (Sec.\ref{Sec2}), followed by a semiclassical quantization of the level \(k\)  (Sec.\ref{Sec3}), assuming that Eq.\ref{51} is given. Next, we outline the hydrodynamic setup (Secs.\ref{Sec4}-\ref{Sec6}) and provide a brief account of the Carter covariant Hamilton principle in hydrodynamics (Sec.VII).

After these preliminaries, we will be ready to address the central part of the paper: the multivalued action (Secs. \ref{Sec7}- \ref{Sec9}). Various forms of the full set of equations of motion are collected in Sec.\ref{Sec10}. The effect of spin introduced by the multivalued action is briefly outlined in Sec.\ref{Sec11} and  Appendix C. The entropy production, the relation to homentropic flow and  comparison with Weyl fermions, is briefly discussed in the Appendices A,B and D.

\section{Chiral Anomaly and the Chiral Phase}\la{Sec2}
Traditionally the chiral anomaly is understood as a linear response of a charged system to the external electromagnetic field. In  the canonical formulation where the particle  current and the momentum are treated as  independent fields the effect of  electromagnetic field is accounted by replacing the kinematic momentum by the canonical momentum 
\be\pi_\mu=p_\mu+A_\mu\,,\la{150}\ee
where \(A_\mu\) is the gauge potential. Then  the helicity  in  the expression for the  Noether current \eq{183}   reads \begin{align}
h^\mu=\varepsilon^{\mu\nu\lambda\sigma}\pi_\nu\p_\lambda
\pi_\sigma\,.\la{174}
\end{align}
Helicity, and therefore the Noether current, is not gauge-invariant.

 Under the gauge transformation   \be \pi_\mu\to  \pi_\mu+\p_\mu{
 \Theta}\ee
it changes as\begin{align}
I^\mu\to I^\mu+k\p_\nu\Theta{}\,^\star\Omega^{\mu\nu}\,,
\end{align} 
 where 
 \be\Omega_{\mu\nu}=\p_\mu
\pi_\nu-\p_\nu \pi_\mu=\p_\mu
p_\nu-\p_\nu p_\mu+F_{\mu\nu}\ee
 is the 4-vorticity tensor and \({}^\star\Omega^{\mu\nu}=\tfrac 12\e^{\mu\nu\lambda\sigma}
\Omega_{\lambda\sigma}\) is the dual tensor. Using the relation  
 \(\p_\mu h^\mu=\tfrac 14\,\Omega_{\mu\nu}{}^\star\Omega^{\mu\nu}\) we write the particle production as  \begin{align}
\p_\mu
n^\mu= - \tfrac k4\,\Omega_{\mu\nu}{}^\star\Omega^{\mu\nu}\,\,.
\label{C}
\end{align}
and transform it in the form of the linear response.   
{ Denote the gauge-invariant part of the Noether current as  
\be j^\mu=n^\mu+\tfrac  k2\Sigma^\mu\,,\la{j}\ee 
where  
\(\Sigma^\mu\) is the {\it kinetic helicity} \cite{abanov2022axial} \begin{align}
\Sigma^\mu=\e^{\mu\nu\lambda\sigma}p_\nu(\p_\lambda
p_\sigma+
F_{\lambda\sigma})\,.\label{Sigma}
\end{align}  
Then the  conservation law \eq{224} or the, equivalently the particle production equation \eq{C} reproduces the commonly
known expression for the  {\it chiral anomaly} 
 \begin{align}
\p_\mu j^\mu={  -}\tfrac k4\,\,F_{\mu\nu}{} ^\star F^{\mu\nu}\,,
\la{12}
\end{align}   
where  \(^\star F^{\mu\nu}=\tfrac 12 \e^{\mu\nu\lambda\sigma}F_{\lambda\sigma} \) is the dual  field tensor. This follows from the identity
\(2\p_\mu\Sigma^\mu=\Omega_{\mu\nu}{} ^\star \Omega^{\mu\nu}-F_{\mu\nu}{} ^\star F^{\mu\nu}\).
}

Two  terms in \(\Sigma^\mu\)  are intrinsically related as they both followed
from \eq{174}. 
{In recent literature, they have been referred to as the {\it chiral vortical effect}
and the {\it chiral magnetic effect}, respectively (see, e.g., \cite{Kharzeev}
and references therein).
The coefficients in (\ref{183},\ref{j},\ref{Sigma},\ref{12}) are quantized topological invariants, as discussed in Sec.\ref{Sec3}. They are  aligned with the result of the direct computation for Weyl fermions of of Ref. \cite{Vilenkin}   as discussed in Appendix D. See, also \cite{FF}. 

We will omit the external gauge field in intermediate formulas of Secs. \ref{Sec4},\ref{Sec5}. The external gauge field could be added upon the use of \eq{150} for the canonical momentum. 
Even at no electromagnetic field   the
canonical momentum and the kinematic momentum momenta are related by a gradient of  a phase \be \pi_\mu=p_\mu+\p_\mu{
 \Theta}\,.\label{P}\ee
  In a usual fluid, the  chiral phase \(\Theta\) has no physical
significance. However, with  the chiral anomaly, the scalar 
\({  \Theta}\) takes on a physical meaning. 
It does not factor into the equation of motion but enters  the
Hamilton functional \eq{37}, and the fluid action, similarly to the {\it
axion} in the theory of CP violation \cite{axion}.
 The chiral phase is an important part of the   of our construct (see, also \cite{Abanov_Cappelli}).

\section{ Semiclassical Quantization of Fluid Helicity, Vortex instantons, and Particle Production}{\la{Sec3}
 While our discussion primarily focuses on classical fluid dynamics, a natural normalization of the fluid helicity arises from Onsager-like semiclassical
consideration.

We recall that in semiclassical fluids  vorticity is localized
in vortex lines (loops in the absence of  spatial boundaries), with    the vortex
circulation \({\cal C}=\oint \bm \pi d\bm x\) being  quantized \cite{Onsager}. Choosing the  Planck constant \(2\pi\hbar
\) as a unit for the momentum   Onsager's quantization states 
\begin{align}
{\cal C}=\oint d\Theta=\text{integer}\,.\la{16}
\end{align} 
The quantization of circulation renders the gauge group
(and thereby the entire fluid phase space) compact. The gauge group becomes
\(U(1)\) making the field \({  \Theta}\) into a phase that winds over
a circle. 

At the same time, the total helicity \be{\cal H}:=\int h^0 d^3\bm x=\int({ \bm
\pi}\cdot\nabla\times \bm \pi)\, d^3\bm x\ee is twice
the linking number of vortex loops in units of vortex circulation  \({\cal H}=2{\rm Lk}[\rm vortex\ loops]\)\, \cite{Moreau,
Moffat}.
In the chosen units it is quantized as an even number \cite{abanov2022axial}.

  Integrating the particle production equation \eq{51} over a time interval and over the entire space we find that the left-hand side of \eq{51} is a change in the total particle number \(\Delta N\) over the time interval. On the right-hand side we obtain the change of the fluid helicity \(\Delta{\cal H}\), i.e. twice the vortex  linking number times \(k/2\)    
 \begin{align}
\Delta N=\tfrac k2\Delta{\cal H}=k\Delta \,\rm Lk[vortex\ loops]\,.\la{15}
\end{align}

Hence,    a   change of the linking of vortex loops
by \( 1\) alters the particle number by   \(k\). Given that the particle number is an integer, \(k\) is also an integer\begin{align}
k\in \mathbb{Z}\,.\la{231}
\end{align}. Also, when momentum is measured in units of the Planck
constant, the particle number current $n^\mu$ and the helicity current $h^\mu$,
the terms in the current \eq{183} and  \eq{C}, should be treated
 of comparable order in gradients.

It follows from \eq{C} that the change of the particle number is assisted
by `vorticity
instantons', a flow which gives a non-zero value to the integral \(\tfrac
14\,\int \Omega\wedge\Omega\), where \(\Omega:=d\pi=\tfrac 12 \Omega_{\mu\nu} dx^\mu\wedge dx^\nu\) is the vorticity 2-form.   This integral is the Pontryagin class, a topological
invariant, of the fluid cotangent bundle. Vorticity instantons, therefore is the flow
  which at an instance  changes the vortex linking number.  

 \section{ currents and  Momentum}  \la{Sec4}

As the particle number is not conserved  \eq{C}, the fluid exchanges particles with a reservoir and, therefore, are  non-homentropic  as we now discuss.    

In relation to chiral fermions, the fluid
could be seen as being composed of particles  with right-handed chirality
(\(k>0\)), and the reservoir represents particles with
the opposite (left-handed)  chirality treated as  a spectator medium (that is a fluid   of massless particles devoid of space-like momentum).

 We denote the particle density number by \(n\) and  the density number of the reservoir  by \(\overline n\) and introduce dimensionless 
 {density ratio} \({ S}= {\overline n}/n\).   Subsequently,
the  fluid energy density \(\varepsilon(n,S)\),  being a function
of \(n\)
is also  a function  of  \({ S}\)  \cite{FootnoteE}. Furthermore, we
assume that  the fluid and the reservoir are oppositely electrically
charged. 
 
 We will consider flows with the density ratio \(S\) varies across streamlines.  Such flow is called non-homentropic. It is also baroclinic.
Non-homentropic flows are endowed with a non-degenerate vorticity 2-form \(\Omega\), and therefore
the top form \(\Omega\wedge
\Omega\), which appears in the particle production equation \eq{C}, is  nowhere
zero (see, Appendix B). This is the essential part of the construction of the multivalued action outlined in Sec. \ref{Sec7} and \ref{Sec8}.    
 
In the
covariant formulation of hydrodynamics \cite{Lichnerowicz,Carter,Gourgoulhon,Markakis}
that we employ here, the equations of motion of the relativistic or non-relativistic
fluid have the same form, although the derivations are technically simpler
in a relativistic setting, which we assume. 

 Taking advantage of the Lorentz metric we express 
the  particle number density
\(n\)  through the particle  current as\begin{align}
n^\mu n_\mu=-n^2\,\label{17}
\end{align}and treat the energy density $\varepsilon$ as a function of \(n^\mu\) and \(S\). We introduce the fluid momentum $p_\mu$ through  a differential of energy taken at a fixed \(S\)
\be
(d\varepsilon)_S=p_\mu dn^\mu
\,.\la{18}
\ee
}   
For isotropic fluid,      
where the energy density depends on \(n\) we express the momentum  in terms of  'enthalpy' per particle   \( w=\p_n\varepsilon\)  and   the 4-velocity  \( u_\mu:=n_\mu\,/n\), a 4-unit  vector collinear to the particle current. Then  
 \begin{align}
p_\mu=w\,  u_\mu,\quad n^\mu=n u^\mu ,\quad u^\mu u_\mu=-1\,.\la{19}
\end{align}
[In  the non-relativistic case,  \eq{19} identifies \(-p_0\) with the energy per particle    \(-p_0=\bm p^2/(2w)+w\)].

Contrary to particles, the constituencies of the reservoir have no momentum and the relation \eq{17} does not hold for the reservoir 
as their density  and their current  are independent.  

\section{Transformations of Natural  Variables}\la{Sec5} Let us examine how the natural variables \(\pi,\Theta,S\) transform under the action of the symmetry  group \(\cal G\). The action of the gauge group is just a variation\begin{align}
{  \Theta}\to {  \Theta}+\delta{  \Theta}\,.
\end{align}The action of the  spacetime diffeomorphisms   \be x^\mu\to x^\mu+\e^\mu(x)\,\label{1944}\ee
  is  carried out by  the Lie derivative \(\cal L_\e\), the directional derivative along a vector field \(\e\).
The  density ratio \(S=\bar n/n\) being a scalar transforms as \begin{align}
\delta_\e {S}:=\mathcal{L}_\e {S}=\e^\mu\p_\mu {S}\,. \quad \label{268}
\end{align}  
  The   momenta \(\pi_\mu\) transform as
the covector defined via a form-valued variation \begin{align}
 \delta_\e \pi:=\mathcal{L}_\e \pi=\delta_\e (\pi_\nu dx^\nu)=(\delta _\e \pi_\nu) dx^\nu\,.\label{267}
\end{align}  
Explicit form of the transformed   momentum      is given by the
{\it Cartan
 formula} followed from \eq{267} \begin{align}
\delta _\e \pi_\nu=\e^\mu\p_\mu \pi_\nu+\pi_\mu\p_\nu\e^\mu
\,.\label{25}
\end{align}

\section{ Hamilton Principle of Hydrodynamics} \la{Sec6}
The Hamiltonian principle asserts that on the equation of motions, the Hamilton functional, which we denote by \(\Lambda\)   is invariant under the action of the group \(\cal G\).  In this form, the Hamiltonian principle incorporates the fluid kinematics into the conservation laws associated with the symmetry group  \(\cal G\) \cite{Arnold}. 

The Hamilton functional depends of {\it natural variables}, which we choose by following Carter \cite{Carter} as the canonical momentum \(\pi\),
the chiral phase \(\Theta \) and the density ratio \(S\). Then  the  variation of the Hamilton functional  
\begin{align}
\delta_\e \Lambda=\int\mathcal{[J^\mu} \delta_\e
\pi_\mu+\pi_S\mathcal{
\delta}_\e { S}-I^\mu\p_\mu\delta{  \Theta}]\,.\la{28}
\end{align} 
defines the conjugate fields:   the flow field \(\mathcal{J}^\mu:=\delta\Lambda/\delta\pi_\mu\),  
a conjugate to the canonical momentum,   the  Noether current \(I^\mu:=-\delta\Lambda/\delta (\p_\mu\Theta)\), 
 and the conjugate
to the density ratio \(\pi_S:=\delta\Lambda/\delta S\).  
 Using explicit forms of the variations
(\ref{268},\ref{25}) a simple algebra leads  to what
 Carter referred  to as the {\it canonical  fluid equation} 
\cite{FootnoteL}
\begin{align}
&\mathcal{J}^\mu\Omega_{\mu\nu}+\pi_\nu (\p_\mu \mathcal{J}^\mu)=\pi_S\,\p_\nu
{ S}\,,\quad \p_\mu I^\mu=0\,.
 \la{1111}
\end{align}
   The first term of the left-hand side  is the force acting
on a rotating fluid parcel.
 It is  balanced by  the force due to the fluid source and the reservoir source. They are  the  second
term sometime called  the  `rocket term' and the  `heat' source on
the right-hand side of \eq{1111}.
A notable feature of the canonical equation is the absence of a reference to a spacetime metric.

The combined result must be gauge-invariant.  The gauge
phase \(\Theta\) introduced through the canonical momentum \eq{P} by the rocket term should not enter the equations. If the flow field \(\cal J\) is gauge-invariant, this is achieved by setting  \(\cal
J\)  divergence free and  killing the rocket term. This is the case of the perfect fluid. However, if \(\cal J\) is not gauge-invariant, i.e., depends on  
    \(\p_\mu\Theta\), the \(\Theta\) in  the  first term must cancel the \(  \Theta\) in the second term. This requirement imposes
a nearly prohibiting condition on \(\cal\ J\).

The perfect fluid could be  defined by the condition that  the flow field, the Noether current and the particle number are all equal \(\mathcal{J}^\mu=I^\mu=n^\mu\). \la{341}
  This  condition determines the Hamilton functional equal to (minus) spacetime integral of the fluid  pressure
 \cite{Whitham,Schutz}\be
\Lambda_0=-\int_{{\cal M}^4}\ P\,.\la{Pressure}
\ee
  It is instructive to check it. For this purpose we need the differential of the fluid pressure with respect to the natural 
variables. The fluid pressure is defined as  \(P= n\p_n \varepsilon-\varepsilon\).
    In   view  of the relations (\ref{17}-\ref{19})  we write the pressure
in terms of momentum and particle current as   \(-P=p_\mu n^\mu+\varepsilon\)
and compute its differential
as\begin{align}
-dP-\p_S\varepsilon\, d S=n^\mu dp_\mu=
&n^\mu d\pi_\mu-n^\mu\p_\mu d{  \Theta}\,.\la{21}
\end{align} It follows that   the flow field and the Noether are equal to the particle number   and   \(\pi_S=\p_S\varepsilon\). We obtain the  canonical form of the Euler equation (also referred as the Lichnerowicz equation) for the  perfect charged 
fluid   \begin{align}
n^\mu\Omega_{\mu\nu}+\pi_\nu(\p_\nu n^\nu)=(\p_\nu\varepsilon)_n\,,\la{31}
\end{align}
where  $(\p_\nu\varepsilon)_n=(\p_S\varepsilon) \,\p_\nu
S$ is the gradient of energy at a fixed \(n\) plus the continuity equation \eq{153}.  The canonical equation is equivalent to the momentum-stress-energy conservation laws (\ref{T},\ref{T1}). the  is simplified by taking into account the continuity equation which kills the rocket term in   in the left-hand side of \eq{31}.
 
The deformation of the Hamilton functional disrupts the accidental relation  between the currents  and alters the mechanism that brings \eq{28} to its gauge-invariant form.
 \\
\section{ Fluid Phase Space and a Generalized Hopf Fibration}\la{Sec7}
The last general  point we need to discuss before introducing the multivalued action is the geometry of the fluid phase space.

 In addition to  the 4-dimensional  space of   momentum,  the phase space  
includes
 the scalar \({ S}\). That makes the phase space  5-dimensional, matching the
dimension of the manifold of the symmetry group \(\cal G\). We illustrate this important feature by invoking the Clebsch realization of the momentum. 

 Vorticity
2-form  \(\Omega=\tfrac
12 \Omega_{\mu\nu}(\dd
x^\mu\wedge
\dd x^\nu)\) of the non-homentropic flow, where \(d{ S}\neq
0\), is non-degenerate \({\rm\ det}\, \Omega_{\mu\nu}\neq 0\). It  endows a
 symplectic structure.   
  Under this condition we may invoke the  Daurboux theorem. It asserts that there are  four local coordinates \(\alpha,\, \beta,\,\eta, S\)  among which one  could be chosen to be the density ratio $S$,
in which  the symplectic structure takes on a canonical form: \be\Omega=d{  \alpha}\wedge d\beta+d\eta\wedge d S\,.\ee
  As  a result,  the canonical momentum is
locally   represented by \underline{five} coordinates   \begin{align}
\pi =d{  \Theta}+{  \alpha} d{  \beta}+\eta\, d S\,.\label{396} \,
\end{align}

 We are endowed with a map of the 5-dimensional phase space, denoted
by \(\mathrm{N}^5\)  to the 4-dimensional spacetime \(\mathrm{M}^4\):
\(\mathrm{N}^5\to \mathrm{M}^4\), where  a point of a  spacetime
\(x\)   is mapped out from a distinct  circle \(\mathrm{S}^1\), represented  by the
 chiral phase \({  \Theta}\).  The  local coordinates of \(\mathrm{N}^5\)  are associated with  five Clebsch potentials \(\Theta,  \alpha, \beta,\eta, S\) \cite{Whitham,Schutz,Lynden-Bell}. Then the canonical momentum  \(\pi=(\p_\mu{  \Theta}+{  \alpha}\p_\mu {  \beta}+\eta\p_\mu  S) dx^\mu\), being the  1-form in \(\mathrm{M}^4\) could be seen as a push-forward of the  1-form \eq{396} 
in \(\mathrm{N}^5\). The map describes  a fibration of  the  phase space
\(\mathrm{S}^1\hookrightarrow \mathrm{N}^5 \to \mathrm{M}^4\), where the spacetime is the base of the bundle. The total space 
\({
\mathrm{N}}^5\)  consists of\(\) fibers, with each fiber being a circle \(\mathrm{S}^1\)
spanned
by the  chiral phase one for each point of the 
spacetime.
This setup is analogous to   the classical Hopf fibration (albeit not for spheres),
 given by the Hopf map \(\mathrm{S}^1\hookrightarrow \mathrm{S}^3\to \mathrm{S}^2\). It was introduced in \cite{Novikov3}.
 The map is characterized by the invariant, which was referred in  \cite{Khesin2} as  Hopf-Novikov
invariant. Analogous to the realization
of the   Hopf invariant in terms of differential forms \cite{Whitehead},
the Hopf-Novikov invariant  is also  represented by  the integral of the  top-form in \(\mathrm{N}^5\), which is
constructed from  the  pull-back of the canonical momentum \cite{Footnote3}
\begin{align}
\mathrm{H}=\int_{\mathrm{N}^5} \pi\wedge (d\pi)^2,\quad (d\pi)^2=d\pi\wedge d\pi \,.\la{33}
\end{align}  
In the
context of semiclassical
hydrodynamics the invariant
is  the volume of the compact phase space.

%

 \section{  Multivalued  Functional}\la{Sec8} The five-dimensional phase space allows the following interpretation. Consider a closed 5-dimensional
space   \(\mathrm{M}^5\) and treat it as a spacetime of  an auxiliary 5-dimensional  fluid.
Then the map  \(\mathrm{M}^5\to \mathrm{N}^5\) defines
the
momentum of the auxiliary fluid via \eq{396} and the invariant \eq{33} is the linking number  of singular 3-surfaces  \begin{align}
\mathrm{H}=\int_{\mathrm{M}^5} \varepsilon^{\mu\nu\lambda\sigma\rho}\pi_\mu
\p_\nu\pi_\lambda\p_\sigma\pi_\rho d^5x \,.\la{331}
\end{align} Consider now  a bounded  5-dimensional
space  \(\mathrm{M}_+^5\), a half-space of \(\mathrm{M}^5\),  and identify   the   boundary of  \(\mathrm{M}_+^5\) with the  physical  spacetime    \( {\mathrm{M}}^4=\p \mathrm{M}_+^5\).    Then the boundary layer  of the 5-dimensional fluid could be identified with the physical fluid. The integrand in 
 \eq{331} is a Jacobian of map \(\mathrm{M}^5\to \mathrm{N}^5\). It is a closed
form \(\pi\wedge (d\pi)^2=d\Phi\).  Therefore, the integral \eq{331} over   \(\mathrm{M}_+^5\) is a
surface
term spanned over physical spacetime, which is the integral over pullback of the 4-form \(\Phi\)
\begin{multline} \mathrm{H}_+=\int_{\mathrm{M}_+^5} \varepsilon^{\mu\nu\lambda\sigma\rho}\pi_\mu
\p_\nu\pi_\lambda\p_\sigma\pi_\rho d^5x=\label{34}
\\
\int_{{\mathrm
M}^4}\Phi \quad ({\rm mod}\, \mathrm H)\,. 
\end{multline} This is Novikov's functional. It is defined 
modulo the invariant
\(\mathrm{H}\) reflecting different choice of \(\mathrm{M}_+^5\).      In this sense the
functional is multivalued. Consequently, the 4-form
 \(\Phi\) can not be expressed in a coordinate-free manner, but it could be elementary computed in chosen coordinates. Choosing the  chiral phase as a fifth coordinate, the density
\(\Phi\), modulo   an exact  4-form, is  \begin{align}
{  \Phi=\tfrac 12{ \Theta}(\Omega\wedge \Omega)}
\,.\label{38}
\end{align} 

 \section { Multivalued  Functional in Fluid dynamics} \la{Sec9}
 Now we  deform the Hamilton functional of the perfect fluid by  the multivalued functional as
   \begin{align}
   \Lambda:=\Lambda_0+\tfrac k4\mathrm{H}_+=
  \int_{M^4}[-P+ \tfrac k8  \Theta(\Omega\wedge \Omega)]\,.\label{37}
\end{align}
While the added functional is not uniquely defined, it nonetheless generates a local
 equation of motion \cite{Novikov2}. The ambiguity of the functional does not extend to its variation as the invariant \(\mathrm H\) does not vary.   
 
 Unlike \(\Lambda_0\),  the functional \(\mathrm{H}_+\) is not gauge-invariant.  It opens a channel of inflow of the 5-dimensional auxiliary fluid
into the physical fluid. Nevertheless,  the equations of motion maintain the gauge invariance. 

We remark that multivalued  term in  \eq{37} is a hydrodynamic version of an {\it
axion}, a ${  \Theta}$-angle promoted to a dynamical field (see \cite{axion}
for a review).

 In the semiclassical fluid, the multivaluedness of \(\mathrm{H}_+\) leads to the quantization of the level \(k\), as we already discussed in Sec.\ref{Sec3}. It follows from the requirement for \(\exp{[(i/\hbar)\Lambda]}\) to be single-valued under a global gauge transformation which changes the circulation by a unit \({\cal C}\to{\cal C}+1\). Then the change  of the   Hamilton  functional  is \(k/2\) times the  of  the  total helicity \(
\Lambda\to\Lambda+\tfrac k2 {\cal H}\). Since the latter is an even integer, \(k\) is quantized (cf., \cite{Witten}).

\section{ Equations of Motion} \la{Sec10} Now we turn to equations of motions.  We calculate the    currents  defined
by \eq{28}, and subsequently,  substitute them into    canonical equation (\ref{1111}). 

First, we     vary \eq{37} over \({  \Theta}\), while keeping 
 \(\pi\) fixed. This gives the divergence of of the Noether current \(\p_\mu I^\mu=\p_\mu
n^\mu+  \tfrac k4\,\Omega_{\mu\nu}{}^\star\Omega^{\mu\nu}\) and  yields  the particle production equation \eq{C}.

The next step is to vary \eq{37} with respect to  \(\pi\), which results in
 a deformation of the flow field \({\cal\ J^\mu}=n^\mu+k\,{}^\star\Omega^{\mu\nu}\p_\nu \Theta\) by a  not gauge-invariant, albeit divergence-free term. The relation \(\p_\mu \mathcal{J}^\mu=\p_\mu n^\mu\)  remains unchanged. 

Now we have all the terms in  the canonical equation \eq{1111}   to verify  that the  chiral phase   \( \Theta\) cancels out. We  see it
with the help of the identity \begin{align}
\e^{\mu\nu\lambda\sigma}X_\rho+\cdots=
0\,,\la{392}
\end{align} which holds for an arbitrary \(X_\mu\) [the ellipsis denotes the cyclic permutation of  five indices] and its consequence \(4(^\star\Omega^{\lambda\mu}\Omega_{\mu\nu})=
\delta^\lambda_\nu(^\star\Omega^{\lambda\mu}\Omega_{\mu\lambda})\). The net result is the familiar canonical form of the Euler equation for a perfect fluid, Eq. (31). The only difference is that the rocket term no longer vanishes 
\begin{align}
n^\mu\Omega_{\mu\nu}+p_\nu (\p_\mu n^\mu)=(\p_\nu\varepsilon)_n \,.\la{46}
\end{align} 
In the form of energy-stress-momentum this equation read  
\begin{align}
\p_\mu \tensor{T}{^\mu_\nu} = F_{\nu\lambda}
n^\lambda,\quad \tensor{T}{^\mu_\nu}=n^\mu p_\nu+\tensor{\delta}{^\mu_\nu}P\,.
  \la{47}
\end{align}
As expected, the Euler equation  is unaffected by the WZN term and remains identical to that of a perfect fluid \cite{FFF}. Since the WZN term is independent of the metric, it does not alter the form of the stress tensor, but only modifies the continuity equation as \begin{align}
 \p_\mu n^\mu= - \tfrac k4\,\Omega_{\mu\nu}{}^\star\Omega^{\mu\nu}\,.
\end{align} These equations are complemented by the equation of state \eq{19} which connects the particle number current and the momentum.
This is the full set. 


\section{ Spin and Spin-Orbit coupling}\la{Sec11}
We conclude by emphasizing a property that the WZN term imparts to the perfect fluid. It gives the fluid  a  spin  equal
to \(k/2\). 

A hint to it 
provides the Newtonian form of the  Euler equation. In this form the  the rocket term in \eq{46}  expressed  in terms of the fluid momentum  by virtue of the  particle production equation is  treated as a force exerted on the fluid. Using \eq{46} and the identity \(\tfrac 12p_\nu(\p_\mu \Sigma^\mu)=\Sigma^\mu\Omega_{\mu\nu}\) followed from \eq{392} we bring the Euler equation into the form  \begin{align}
n^\mu\p_\mu p_\nu+\p_\nu P=k\Sigma^\mu\Omega_{\mu\nu} \,.\la{48}
\end{align}

The term on the right-hand side of \eq{48} indicates that our fluid is spinning, with the spin density  \(\tfrac 12\Sigma^\mu\) (see, Appendix C). The spin exerts a force  on fluid vorticity given by the right-hand side of \eq{48}   (more about the spin density and the spin current is given in Appendix D).\bigskip

In summary, we presented what we believe to be the only deformation of a single-component perfect fluid that does not introduce additional scales into the system. This deformation captures the chiral anomaly. While it modifies the continuity equation, it does not alter the form of the stress-energy tensor. It requires the system to have an open channel for particle production and breaks inversion symmetry in the same manner as the chiral anomaly. The kinematic and geometric properties of our hydrodynamics are consistent with those of systems composed of Weyl (or chiral) fermions.

\begin{acknowledgments}The author
gratefully acknowledges discussions with
 V. P. Nair,
  G. Volovik,
 and  M. Stephanov,
 L. Friedlander. Special appreciation is extended to A. G. Abanov and A.
Cappelli for their collaboration on this subject. The work was supported by the NSF under Grant NSF
DMR-1949963. 
\end{acknowledgments}

\section*{Appendix A.\quad Vorticity instantons and `entropy' production}\la{Sec1} Here  we remark  on the role of the anomaly in particle exchange with the  reservoir. 

We recall that the reservoir density \(\bar n=nS\) could be  interpretation as an `entropy' \cite{FootnoteE}. Consequently,  the `entropy current'   is \(s^\mu:=\bar n u^\mu=S n^\mu\).    By contracting  \eq{46} with \(n^\mu\) we obtain the   relation between  particles and `entropy' productions
\(\p_\mu s^\mu=-(\p\bar n/\p n)_\varepsilon (\p_\mu n^\mu)\), with \(\bar n \) being treated as a function of \(\varepsilon\) and \(n\).  This is a general relation for an open system. It means   that the entropy propagates along isoenergy hypersurfaces $d\varepsilon=0$.  Specifically to our fluid we may express the entropy production in terms of vorticity as
\begin{align}
\p_\mu
s^\mu=  \frac k4\,\left(\frac{\p\bar n}{\p n}\right)_\varepsilon \Omega_{\mu\nu}{}^\star\Omega^{\mu\nu}\,.
\end{align}  
We conclude that the entropy production goes along with particle production assisted by vorticity instantons. A change of  linking number of vortex loops changes the entropy \cite{F4}.
%
%

\section*{Appendix B.\quad Homentropic Flows}A homentropic flow occurs when the density ratio \( S\)  is uniform and constant. It is also called uniformly canonical \cite{Carter}. In this situation, the flow is   barotropic and the vorticity tensor is degenerate \({\rm\ det}\,\Omega_{\mu\nu}=0\), having rank 2.   Consequently,   
 the rate of particle production  \(^\star\Omega^{\mu\nu}\Omega_{\mu\nu}=\sqrt{{\rm\ det}\,\Omega_{\mu\nu}} \) vanishes.  This prevents the construction of the WZN-term since the phase
space of a homentropic flow is not symplectic. In this case canonical  helicity and particle currents are conserved independently \(\p_\mu n^\mu=\p_\mu h^\mu=0\).  In this case, the equations of motion
are no different from that of the perfect fluid but the kinematic  helicity   \(\Sigma^\mu\) obeys the anomaly equation  \(\p_\mu \Sigma^\mu={  -}\tfrac 12\,\,F_{\mu\nu}{} ^\star F^{\mu\nu}\) 
\cite{AMN,abanov2022axial,wiegmann2022chiral,AW2,wiegmann2023Hamilton,Abanov_Cappelli}. \\

\section*{Appendix C.\quad  Totally antisymmetric spin current}
A spinning fluid possesses a spin current, which we  denote by $\tensor{\sigma}{_\mu_\alpha_\beta}$. 
Given spin current one finds a spin tensor  $\sigma_{\alpha\beta}=- u^\nu\tensor{\sigma}{_\nu_\alpha_\beta}$ by projecting of the spin current onto the 4-velocity. It is customary  to use the dual spin tensor $^\star{}\tensor{\sigma}{^\mu^\nu}=\tfrac 12\e^{\mu\nu\alpha\beta}\tensor{\sigma}{_\alpha_\beta}$. Then the spin density reads $\sigma^\nu=u_\mu \tensor{^\star \sigma}{^\mu^\nu}$. 

We identify the spin density as a neutral part of the current \eq{j}. It is given by the half of helicity  as Eq.\eq{Sigma} suggests   \be \sigma^\mu=\tfrac 12\Sigma^\mu,\quad \tensor{\sigma}{_\mu_\alpha_\beta}=-\tfrac 16p_{[\mu}(\p_\alpha p_{\beta]}+F_{\alpha\beta]})\,,\ee
 Here $[\mu,\alpha, \beta]$ denotes the antisymmetrization over all three indices. We observe that the spin current is totally antisymmetric. This is a distinguished property of spinning fermions.  

\section*{Appendix D.\quad  Comparison with the kinematics of Weyl fermions}

Our results are consistent with the 
 kinematics of  chiral  fermions. Here we briefly outline the major points.  

Chiral fermions carry an electric charge and also  \( \tfrac 12\)-spin.  The current, therefore  is composed of a charge  (the particle number) current  and a neutral part representing the spin current. The formula \eq{j} represents this composition.  Therefore, we identify the neutral  part of the current  $\sigma^\nu=\tfrac 12 \Sigma^\nu$ withe the  spin.  The neutral part  can be also obtained as a difference between the  particle current and  the current of antiparticles    \(\Sigma^\nu= j^\nu |_{\mu}- j ^\nu|_{-\mu}\), where $\mu$ is the chemical potential. The (twice) of the spin current of free chiral fermions had been computed  by Vilenkin in a series of papers \cite{Vilenkin}.  Despite that Vilenkin computations valid for free fermions, some of the results are transferred to a liquid state and could be compared with formulas (\ref{j},\ref{Sigma}).

  Vilenkin computed the
 equilibrium value of the space-like component of    $\bm \Sigma$ for
free rotating chiral fermions  in a magnetic
field and at a low temperature. His  result  (in units of Planck constant $2\pi\hbar$) is 
\be \bm\Sigma=2[(\mu^2+\tfrac {\pi^2T^2}{3} )\bm \omega+\mu\bm B]\,.\la{V}\ee
Now we specify our formula  \eq{Sigma} 
 \(\Sigma^\mu=\e^{\mu\nu\lambda\sigma}p_\nu(\p_\lambda
p_\sigma+
F_{\lambda\sigma})\) for a   stationary  rotating fluid and check  it against Vilenkin's direct computation.
 
  The space-like component of  \(\Sigma\) evaluated for a stationary flow  in the leading  order of velocity \(\bm u\) is \begin{align}
\bm\Sigma=w^2\nabla\!\times\!\bm u-2p_0
\bm B\,.
\end{align}
%
%
%
%
%
 We can
extend this formula to a rotating fluid, by adding   the twice the angular velocity of rotation \(2 \bm\omega\)  to vorticity  \(\bm\Sigma=w^2(\nabla\!\times\bm u+2\bm\omega)-2p_0\bm B\).  At equilibrium we drop vorticity.  Then  \begin{align}
\bm\Sigma=2(w^2\bm\omega-p_0\bm
B)\,.\la{53}
\end{align} 
This is a general result. To specify it for fermions we need the value of enthalpy for a degenerate  Fermi gas.  It could  be extracted from the textbook \cite{LL}, par.61, problem 2. At  low temperature the enthalpy of free fermions is \(w\approx\mu(1 +\frac {\pi^2}{6}\tfrac{T^2}{\mu^2})\). For the Fermi gas we have to set $-p_0$ to be the Fermi level equal  the chemical potential $-p_0=\mu$. Combining, we get Vilenkin's result \cite{F5}.
%


\begin{thebibliography}{10}
\bibitem{FootnoteE} In a nonhomentropic flow, the fluid energy depends on the particle
density \(n\) and another variable  commonly associated
with the entropy. Following Carter \cite{Carter}
 the entropy'  could be identified with a  particle density of the reservoir
 \(\overline n\),  and the density ratio \(S\) with a specific entropy. Then the partial derivatives  \(w=\p_n\varepsilon\) and \(T=\p_{\bar n}\varepsilon=n^{-1}\p_S\varepsilon\) are  the analogs of thermodynamics specific enthalpy and temperature. The latter should not be confused with a physical temperature, which we consider to be zero.
\bibitem{FootnoteL}This important form of fluid mechanics is due to Lichnerowicz
\cite{Lichnerowicz}, and was obtained  from the Hamilton principle  by Carter \cite{Carter}.
\bibitem{F5} Notice that, in the liquid  state $-p_0=wu_0\approx \mu+TS$, yielding a sub-leading correction  to the magnetic field  dependence in the spin current of free fermions (the last term of \eq{V}). This issue was pointed out by the anonymous referee.
\end{thebibliography}
\end{document}